\documentclass[12pt]{article}

\begin{document}

\def\be{\begin{equation}}
\def\ee{\end{equation}}
\def\bea{\begin{eqnarray}}
\def\eea{\end{eqnarray}}
\def\nn{\nonumber \\}
\def\e{{\rm e}}

\thispagestyle{empty}

\  \hfill
\begin{minipage}{3.5cm}
YITP-02-40 \\
July 2002 \\
%hep-th/0207009 \\
\end{minipage}

\vfill

\begin{center}

{\large\bf The AdS/CFT Correspondence and Logarithmic Corrections
to Braneworld Cosmology and the Cardy-Verlinde Formula}

\vfill

{\sc James E. Lidsey$^\heartsuit$}\footnote{email: J.E.Lidsey@qmul.ac.uk},
{\sc Shin'ichi Nojiri$^\spadesuit$}\footnote{
email: snojiri@yukawa.kyoto-u.ac.jp,
nojiri@nda.ac.jp}, \\
{\sc Sergei D. Odintsov$^\clubsuit$}\footnote{
email: odintsov@mail.tomsknet.ru} and
{\sc Sachiko Ogushi$^\diamondsuit$}\footnote{JSPS fellow,
email: ogushi@yukawa.kyoto-u.ac.jp}

\

Astronomy Unit, School of Mathematical Sciences, Queen Mary \\
University of London, Mile End Road, LONDON, E1 4NS, U.K.

\

$\spadesuit$ Department of Applied Physics,
National Defence Academy \\
Hashirimizu Yokosuka 239, JAPAN

\

$\clubsuit$ Lab. for Fundamental Study \\
Tomsk Pedagogical University, 634041 Tomsk, RUSSIA \\

\

$\diamondsuit$ Yukawa Institute for Theoretical Physics,
Kyoto University, Kyoto 606-8502, JAPAN \\

\vfill

{\bf ABSTRACT}

\end{center}

\noindent The AdS/CFT correspondence is employed to derive logarithmic
corrections to the Cardy-Verlinde formula when
thermal fluctuations in the Anti-de Sitter black hole are
accounted for. The qualitative effect of these corrections on
the braneworld cosmology is investigated.
The role of such terms in enabling a contracting
universe to undergo a bounce is demonstrated.
Their influence on the stability of black
holes in AdS space and the Hawking-Page-Witten phase transitions is also
discussed.

\noindent
PACS: 98.80.Hw, 04.50.+h, 11.10.Kk, 11.10.Wx

\newpage

\section{Introduction}

The holographic principle implies that the number of degrees of freedom
associated with gravitational dynamics is determined by the boundary of the
spacetime rather than by its bulk \cite{holographic}.
The anti--de Sitter/conformal field theory
(AdS/CFT) correspondence represents a realization of this principle by
providing a duality between classical $d$--dimensional gravity in AdS space
and a quantum CFT located on its boundary \cite{adscft,witten,ads1}.
A further connection between higher--dimensional classical gravity and
lower--dimensional quantum physics was uncovered by
Verlinde \cite{EV}, who showed that
the standard Friedmann--Robertson--Walker (FRW)
equations for a (spatially closed) radiation dominated universe
can be rewritten in a form that is
equivalent to the Cardy \cite{cardy} formula for the
entropy of a two--dimensional CFT.

Insight into the origin of such a Cardy--Verlinde formula can be gained
from the braneworld scenario.
It has recently become clear that braneworld cosmology is closely
related to the AdS/CFT correspondence and, in particular, the
Randall--Sundrum
braneworld \cite{rs} may be viewed as a specific manifestation of this
framework, where the CFT--dominated universe is interpreted as
a co--dimension one brane representing the (dynamical) boundary
of Schwarzschild--Anti--de Sitter (SAdS) space \cite{hhr,SV}.
The motion of the brane away from the black hole is interpreted by an
observer on the brane in terms of Hubble expansion and the classical
dynamics
of the expansion is formally equivalent to that of a
radiation--dominated FRW universe \cite{branedynamics}.
Following Witten's identification of
the entropy of the AdS black hole with the entropy of the
dual CFT \cite{witten}, it can be shown that the FRW equations
correspond to the Cardy entropy formula of the CFT
when the brane passes through the black hole event
horizon\footnote{Of course,
the precise relation is dependent on the form of bulk space
and, in particular, whether it is a pure or
asymptotically AdS space, an AdS black hole, or even de Sitter
space. In practice, each possibility should be
considered independently.} \cite{SV}.
Furthermore, such a form for the FRW equation defines a new dynamical bound
on the cosmological entropy \cite{EV}.
(For a recent review and list of references, see \cite{NOO}).

In the present letter, we take into account corrections to the entropy
of the five--dimensional AdS black hole that arise due
to thermal fluctuations around its equilibrium state.
It has been known for some time that such fluctuations
result in logarithmic corrections to the black hole entropy
\cite{log6,log7,log8,log8.5,log9,log10,log11,Log} and this is
also the case for AdS black holes. Previous studies of the CV formula
(or the corresponding FRW equation and cosmological
entropy bound) in an AdS/CFT context have thus far
neglected this sub--dominant, but important,
contribution.  In this letter, it is shown that
these effects result in logarithmic corrections to the
CV formula and the brane FRW equations.
Moreover, the Hawking--Page phase transitions \cite{hp},
interpreted by Witten as confinement--deconfinement transitions
in the dual CFT \cite{witten}, may also be influenced by these corrections.
The modified CV formula is derived in Section 2 and we proceed
to investigate the cosmological implications of the logarithmic
corrections in Section 3. We conclude with a discussion in Section 4.

\section{Logarithmic Corrections to the
Cardy--Verlinde Formula}

The metric of the
five--dimensional SAdS black hole is
given by
\bea
\label{SAdS}
ds^{2}_{5} &=& -\e^{2\rho} dt^2 + \e^{-2\rho} da^2
+ a^2 \sum_{ij}^{3}g_{ij}dx^{i}dx^{j} \ ,\nn
\e^{2\rho}&=&{1 \over a^{2} }\left( -\mu + {k\over 2}a^{2}
+ {a^4 \over l^2} \right) ,
\eea
where
$l$ represents the curvature radius of the SAdS bulk space.
The three--metric, $g_{ij}$,
is the metric of an Einstein space with Ricci tensor given by
$r_{ij} =kg_{ij}$, where the constant $k=\{ -2, 0, +2 \}$
in our conventions.
The
induced metric on the brane is then negatively-- curved, spatially flat,
or positively--curved depending on the sign of $k$.

The mass of the
black hole is parametrized by the constant, $\mu$, and may be expressed
directly in terms of the horizon radius, $a_H$, of the black hole:
\bea
\label{ab00}
\mu =a_{H}^{2} \left( {a_{H}^{2} \over l^{2}}
+ {k \over 2} \right) ,
\eea
where
\bea
\label{abrh1}
a_{H}^{2}=-{kl^{2} \over 2} + {1\over 2}
\sqrt{ l^{4}+ 4 \mu l^{2} }
\eea
is the largest solution to the constraint equation
$\exp[2\rho (a_H)]=0$.

The CV formula is derived from the thermodynamical properties
of the five--dimensional black hole.
The free energy, $F$, the entropy, ${\cal S}$,
the thermodynamical energy, $E$, and the
Hawking temperature, $T_H$, of the black hole are
given, respectively, by
\bea
\label{ab3}
F &=& -{W_3 \over 16\pi G_5} a_{H}^2
\left( {a_{H}^2\over l^2} - {k \over 2}  \right)\; ,\quad
{\cal S }= {W_{3} a_H^3 \over 4 G_5} \; ,\\
\label{ab4}
E&=& F + T_{H} {\cal S} = {3W_{3}\mu \over  16 \pi G_5 }
, \quad
T_H = {k \over 4\pi a_{H}} +{a_{H} \over \pi l^2}  ,
\eea
where $W_3$ is the volume of the unit three--sphere and $G_5$ denotes
the five--dimensional Newton constant.

In this paper, we are interested primarily in the
corrections to the entropy (\ref{ab3}) that
arise due to thermal fluctuations.
An expression for the leading--order correction
has been found for a generic thermodynamic system \cite{log11}.
The entropy is calculated in terms of a grand
canonical ensemble, where the corresponding
density of states, $\rho$, is determined
by performing an inverse Laplace transformation
of the partition function\footnote{The reader is referred to Refs.
\cite{log11,Log}
for details.}. The integral that arises in this procedure
is then evaluated in an appropriate saddle--point
approximation. The correction to the entropy follows by assuming that
the scale, $\epsilon$, defined such that ${\cal{S}} \equiv
\ln (\epsilon \rho )$, varies in direct proportion
to the temperature, since this latter parameter is the only
parameter that provides a physical measure of scale in the
canonical ensemble.
The final result is then of the form \cite{Log}:
\begin{equation}
\label{gencorr}
{\cal{S}} = {\cal{S}}_0 -\frac{1}{2} \ln C_v + \ldots  ,
\end{equation}
where $C_v$ is the specific heat of the system evaluated at
constant volume and ${\cal S}_0$ represents the uncorrected entropy.
In the case of the AdS black hole, this is given by
Eq. (\ref{ab3}). The specific heat of the black hole is determined
in terms of this entropy:
\bea
C_{v} \equiv \left( {3 W_3 \over 16\pi G_5}\right){d\mu \over dT_H}
= 3 {2a_H^2+{k\over 2}l^2 \over 2a_H^2 -{k\over 2} l^2}{\cal S}_0
\eea
and for consistency, the condition
$a_{H} > l^2 k /4$ must be satisfied
to ensure that the specific heat is positive. In this limit, $C_v
\approx 3 {\cal{S}}_0$, and this implies that \cite{Log}
\bea
\label{total}
{\cal S} = {\cal S}_0
-{1\over 2} \ln {\cal S}_0+ \cdots  \; .
\eea

Given the form of the logarithmic correction (\ref{total}) to the entropy,
it is now possible to derive the corresponding corrections
to the CV formula.
We begin by recalling that
the four--dimensional energy, derivable from the FRW equation of motion
for a brane propagating in a SAdS background, is given by
\bea
\label{e444}
E_{4} &=& {3 W_{3} l \mu \over  16 \pi G_5 a}
\eea
and is related to the five--dimensional energy
(\ref{ab4}) of the bulk black hole such that
$E_{4} =(l/a) E$ \cite{SV}. This implies that the temperature,
$T$, associated with the brane should
differ from the Hawking temperature (\ref{ab4})
by a similar factor \cite{SV}:
\be
\label{abe22}
T={l \over a}T_H
={a_H \over \pi a l} + {l k \over 4\pi a a_H} .
\ee

In determining the corrections to the entropy, a crucial
physical parameter is the Casimir energy, $E_C$ \cite{EV}, defined in terms
of the
four--dimensional energy, $E_4$, pressure, $p$, volume, $W$,
temperature, $T$, and entropy, ${\cal{S}}$:
\be
\label{abEC1}
E_C=3\left( E_4 + pW - T{\cal S}\right)\ .
\ee
This quantity vanishes in the
special case where
the energy and entropy are purely extensive,
but in general,
this condition does not hold.
For the present discussion,
the total entropy is assumed to be of the form
(\ref{total}),
where the uncorrected entropy, ${\cal{S}}_0$,
corresponds to that associated with the black hole
in Eq. (\ref{ab3}) (due to the AdS/CFT correspondence). It then follows by
employing
(\ref{e444}) and (\ref{abe22})
that the Casimir energy (\ref{abEC1})
can be expressed directly in terms of the uncorrected entropy:
\be
\label{abEC2}
E_C= {3 l a_H^2 W_3 k \over 16 \pi G_5 a}+{3\over 2}  T \ln {\cal S}_0 \ ,
\ee
where the direct dependence on the pressure has been
eliminated by means of the relation $p=E_4/(3W)$.

Moreover, in the limit where the logarithmic correction in Eq.
(\ref{abEC2})  is small, it can be shown, after substitution
of Eqs. (\ref{ab3}), (\ref{e444}), and
(\ref{abEC2}), that the four--dimensional and Casimir
energies are related to the uncorrected entropy by
\cite{NOO}
\be
\label{abSS}
{4\pi a \over 3 \sqrt{|k|} }\sqrt{\left|
E_C\left( E_{4} - {1\over 2} E_C\right)\right|}
\sim
{\cal S}_0 + {\pi a l \over k a_H^3} T
\left( {a_H^4 \over l^2} - {k\over 2} a_H^2
\right) \ln {\cal S}_0 .
\ee

The coefficient of the logarithmic term on the
right--hand side of
Eq. (\ref{abSS}) is constant. In the limit where
the correction is small, this constant can be
expressed directly in terms of the four--dimensional and
Casimir energies
through the relationship:
\be
\label{SN2}
{\pi a l \over k a_H^3} T \left( {a_H^4 \over l^2} - {k\over 2}
a_H^2 \right)
= {\left( 4 E_4 -E_C \right) \left( E_4 -E_C \right)
\over 2 \left( 2E_4 -E_C \right) E_C }  ,
\ee
where we have substituted in Eq. (\ref{abe22}) for the temperature and
have also employed the relation
\be
\label{SN1}
{E_4 - {1 \over 2}E_C \over E_C}={a_H^2 \over kl^2}\ .
\ee

We may conclude, therefore, that in the limit where the logarithmic
corrections are sub--dominant, Eq. (\ref{abSS})
can be rewritten to express the entropy directly in terms
of the four--dimensional and Casimir energies (corrected Cardy-Verlinde
formula \cite{NOO}):
\bea
\label{SN3}
{\cal S}_0 =
{4\pi a \over 3 \sqrt{|k|} }\sqrt{\left| E_C\left( E_{4}
- {1\over 2} E_C\right)\right|} \nn
-{\left( 4 E_4 -E_C \right) \left( E_4 -E_C \right)
\over 2 \left( 2E_4 -E_C \right) E_C }
\ln \left({4\pi a \over 3 \sqrt{|k|} }\sqrt{\left| E_C\left( E_{4}
 - {1\over 2} E_C\right)\right|}\right)
\eea
and, consequently, the total entropy Eq. (\ref{total}),
to first--order in the logarithmic term,
is given by \cite{NOO}
\bea
\label{ttol}
{\cal S}
\simeq  {4\pi a \over 3 \sqrt{|k|}} \sqrt{\left| E_C\left( E_{4}
 - {1\over 2} E_C\right)\right|}\nn
- {E_4 \left( 4E_4 -3E_C \right) \over 2\left( 2E_4 -E_C \right) E_C }
\ln \left({4\pi a \over 3 \sqrt{|k|} }\sqrt{\left| E_C\left( E_{4}
 - {1\over 2} E_C\right)\right|}\right)\; .
\eea
It then follows that
the logarithmic corrections to CV formula
are given by the second term on right--hand--side of Eq. (\ref{ttol})
and the magnitude of this correction
can be deduced by taking the logarithm
of the original CV formula. As we saw in above discussion these corrections
are caused by thermal fluctuations of the AdS black hole.

Finally, the four--dimensional FRW equation
also receives corrections
as a direct consequence of the logarithmic correction
arising in Eq. (\ref{ttol}). In general, the Hubble
parameter, $H$, is related to the four--dimensional (Hubble)
entropy by \cite{SV}
\bea
\label{HS}
H^2 = \left({2 G_4 \over W }\right)^2 {\cal S}^2\; ,
\eea
where the effective four--dimensional Newton constant,
$G_4$, is related to the five--dimensional Newton constant,
$G_5$, by $G_4=2G_5/l$. Hence, by substituting
Eq. (\ref{ttol}) into Eq. (\ref{HS}), it can be
shown by employing
Eqs. (\ref{ab00}), (\ref{ab3}), (\ref{abe22}),
(\ref{abEC2}), (\ref{SN3}) and (\ref{ttol}),
that the four--dimensional FRW equation is given by
\bea
\label{lnln}
H^2 &=& \left({2 G_4 \over W }\right)^2 \left[
\left( {4\pi a \over 3 \sqrt{|k|}} \right)^2
\left| E_C\left( E_{4} - {1\over 2} E_C\right)\right| \right.
-{4\pi a \over 3 \sqrt{|k|}}
{E_4 \left( 4E_4 -3E_C \right) \over \left( 2E_4 -E_C \right) E_C }\nn
&& \times \left.\sqrt{\left| E_C\left( E_{4}
 - {1\over 2} E_C\right)\right|}
\ln \left({4\pi a \over 3 \sqrt{|k|} }\sqrt{\left| E_C\left( E_{4}
 - {1\over 2} E_C\right)\right|}\right)\ \right] \nn
&=& -{k \over 2 a_H^2} +{8\pi G_4 \over 3} \rho
-{2G_4 \over Wl} \ln {\cal S}_0\;,
\eea
where logarithmic corrections have been included
up to first--order in the logarithmic term,
the effective energy density
is defined by $\rho=E_4/W$ and
$W=a_H^3 W_3$ parametrizes the spatial volume of the
world--volume of the brane.
Since the first term on
the right--hand--side of Eq. (\ref{lnln}) is identical to the standard
(radiation--dominated) FRW equation in the limit where the
scale factor, $a$, of the brane coincides with the horizon radius,
$a_H$, of the black hole,
the logarithmic corrections for
the FRW equation are given by the second term on the
right--hand--side in terms of the uncorrected entropy
(\ref{ab3}) of the black hole.

\section{Qualitative Dynamics of the Brane Cosmology}

In this section we investigate the
asymptotic behaviour of the FRW brane cosmology
when the logarithmic corrections to the CV formula
are included. Formally, the FRW equation
(\ref{lnln}) holds precisely at the instant when the brane crosses the black
hole horizon. Here we extend the analysis to consider
an arbitrary scale factor, $a$, where the world--volume of the brane
is given by the line--element
$ds^2_4 = -d\tau^2 +a^2(\tau)
g_{ij}dx^{i}dx^{j}$. Thus,
the FRW equation is given by
\bea
\label{friedmannlog}
H^2 = -{k \over 2 a^2} +{8\pi G_4 \over 3} \rho
-{2 G_4 \over W l} \ln {\cal S}_0\;,
\eea
where $W=a^3 W_3$ and ${\cal S}_0 =W_3 a^3 /(4G_5)$.
Eq. (\ref{friedmannlog}) can be rewritten in such a way that
it represents the conservation
of energy of a point particle moving in a
one--dimensional effective potential, $V(a)$:
\bea
\left( {da \over d\tau }\right)^2 &=& -{k \over 2} -V(a) \\
\label{effec}
V(a) &\equiv& -{8\pi G_4 \over 3}a^2 \rho + {2G_4 a^2 \over Wl}
\ln {\cal S}_0\ ,
\eea
where, in this interpretation, the variable $a$ represents the
position of the particle.
Since $\rho \propto a^{-4}$, the first term in
the effective potential (\ref{effec}) redshifts as
$a^{-2}$ as the brane moves away from the black hole horizon.
This term is often referred to as the `dark radiation' term.

To proceed, let us briefly
recall the behaviour of the standard FRW cosmology, whose effective
potential includes only the first term on the right--hand side
of Eq. (\ref{effec}). The behaviour of this
potential is illustrated in Fig. 1. The brane
exists only in regions where the line $V(a) \le -k/2$
(so that $H^2 > 0$).  For the case of $k=2$,
the spherical (de-Sitter) brane expands from
an initial state at $a=0$ and
reaches a maximal size $a_{\rm max}$ before
re-collapsing.  On the other hand,
the brane expands to infinity for the spatially
flat $(k=0)$ and hyperbolic $(k=-2)$ geometries.

\begin{figure}[htbp]
\begin{center}
\unitlength=0.47mm
\begin{picture}(160,110)
\thicklines
\put(30,93){$V(a)$}
\put(152,58){$a$}
\put(40,60){\vector(1,0){110}}
\put(40,10){\vector(0,1){80}}
\qbezier[400](42,10)(45,40)(60,50)
\qbezier[400](60,50)(72,58)(150,59)
\put(0,68){$-{k \over 2}=1$}
\put(0,58){$-{k \over 2}=0$}
\put(0,48){$-{k \over 2}=-1$}
%\put(73,57){$a=a_{\rm max}$}
\put(52,62){$a=a_{\rm max}$}
\thinlines
\put(40,70){\line(1,0){110}}
\put(40,50){\line(1,0){110}}
\put(60,50){\line(0,1){10}}
\end{picture}
\end{center}
\caption{The standard effective potential for the evolution
of FRW universe.  For the case of $k=2$,
the spherical (de-Sitter) brane starts from $a=0$ and
reaches its maximal size $a_{\rm max}$ and then it
re-collapses.}
\end{figure}
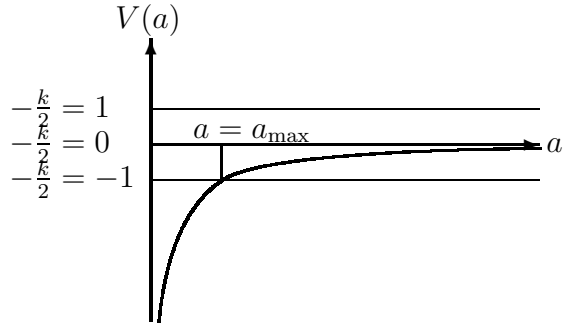

\begin{figure}[htbp]
\begin{center}
\unitlength=0.47mm
\begin{picture}(160,110)
\thicklines
\put(30,93){$V(a)$}
\put(152,58){$a$}
\put(40,60){\vector(1,0){110}}
\put(40,10){\vector(0,1){80}}
\qbezier[400](41,10)(42,82)(70,75)
\qbezier[400](70,75)(102,67)(150,65)
\put(0,68){$-{k \over 2}=1$}
\put(0,58){$-{k \over 2}=0$}
\put(0,48){$-{k \over 2}=-1$}
%\put(73,57){$a=a_{\rm max}$}
%\put(52,62){$a=a_{\rm max}$}

\thinlines
\put(40,70){\line(1,0){110}}
\put(40,50){\line(1,0){110}}
\qbezier[400](41,10)(42,72)(70,65)
\qbezier[400](70,65)(82,62)(150,61)
\end{picture}
\end{center}
\caption{The effective potential (\ref{effec}) with logarithmic corrections
included.
The continuous line represents the potential
for the case where it crosses the line
$V(a)=1$ at a finite value of the scale factor.
The dotted line corresponds to the region of parameter
space where $V(a) < 1$ for all values of the scale factor.
As we discuss in the text, the dotted potential
arises if $l_P/l \ll 1$, which is the limit expected from the
point of view of the AdS/CFT correspondence.}
\end{figure}
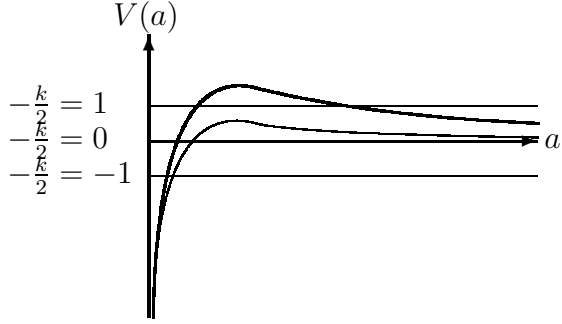

The logarithmic correction to the
effective potential (\ref{effec})
is proportional to $(\ln a)/a$ and
the corrected form of the potential
is shown in Fig. 2.
The asymptotic behaviour of the brane is now as follows:
\begin{itemize}
\item $k=2$: The behaviour of the
positively curved brane is qualitatively similar to that of the
uncorrected scenario. The brane expands from an initial state of
vanishing spatial volume, $a=0$, and
reaches a point of maximal expansion, $a_{\rm max}$, where this latter
quantity is defined by the constraint equation
$V(a_{\rm max})=-1$ in Eq. (\ref{effec}). Re-collapse to zero volume
then ensues.
\item $k=0$: As in the closed case, the spatially
flat brane has vanishing spatial volume initially. However,
the late--time behaviour of this model is radically
altered by the logarithmic correction term.
The brane reaches a maximal size, $a_{\rm max}$,
where $V(a_{\rm max})=0$, and then undergoes a recollapse.
This change in behaviour arises as a direct result
of the logarithmic term.
\item $k=-2$:
There are a number of possible outcomes for the hyperbolic
brane depending on the choice of parameters. In the case where
the condition $V(a) <1$ is always satisfied, the brane expands
to infinity. On the other hand, if the potential
exceeds the critical value of unity, the brane reverses its direction
of motion through the bulk space.
Moreover, the constraint
equation $V(a)=1$ admits two solutions
$a_1, a_2,\; (a_1 < a_2)$. The brane may either
expand from $a=0$ to
reach a maximal size, $a_1$, before re-collapsing, or
alternatively, it may initially have infinite spatial volume
and undergo a collapse to a minimal size,
$a_2$, re-expanding to infinity. This latter behaviour
is an example of a `bouncing' cosmology.
Such behaviour is not possible within the
context of the standard FRW equation.
\end{itemize}

We now proceed to examine the qualitative behaviour outlined above
in more detail.
It proves convenient to define a rescaled scale factor
\begin{equation}
b \equiv \left( \frac{W_3}{2lG_4} \right)^{1/3} a
\end{equation}
and rescaled parameters
\begin{equation}
\label{tilde}
\tilde{k} =\frac{k}{2} \left( \frac{W_3}{2lG_4} \right)^{2/3}
, \qquad \tilde{\mu} =
\left( \frac{W_3}{2lG_4} \right)^{4/3} \mu  .
\end{equation}
The Friedmann equation (\ref{friedmannlog}) may then be
expressed in the form
\begin{equation}
\frac{\dot{b}^2}{b^2} = -\frac{\tilde{k}}{b^2} +\frac{\tilde{\mu}}{b^4}
-\frac{3}{l^2} \frac{\ln b}{b^3}  ,
\end{equation}
where we have employed Eq. (\ref{e444}) to re--introduce the black hole
mass parameter, $\mu$.

The qualitative dynamics of the brane can be determined in terms
of the function $f\equiv - b\ln b$ by a further rewriting
of the Friedmann equation to
\begin{equation}
\label{friedmannf}
H^2 = \frac{1}{b^4} \left[ \tilde{\mu} -\tilde{k} b^2 +
\frac{3}{l^2} f (b) \right]  .
\end{equation}
The first and third terms on the right--hand--side
of Eq. (\ref{friedmannf}) are
semi positive--definite in the region
$0\le b \le 1$. The importance of the function $f(b)$ is that it becomes
negative for $b>1$. Thus, a recollapse of the brane may
ensue if the magnitude of the third term of the right hand side of
Eq. (\ref{friedmannf}) comes to dominate. This function
exhibits only one turning point, a maximum at
$b=e^{-1}$, such that
$f_{\rm max} =e^{-1}$ and this implies that $f \rightarrow 0$
as $b \rightarrow 0$. Thus,
as the scale factor
tends to zero, the logarithmic correction dominates the curvature term, but
both these terms are themselves dominated by the dark radiation
term. Consequently, the asymptotic behaviour of the scale
factor in the limit of small spatial volume
corresponds to that of a
radiation/CFT--dominated universe, $a \propto t^{1/2}$.

Let us now focus on the spatially flat model, $k=0$.
Since $0 \le f \le e^{-1}$ for $0 \le b \le 1$, the logarithmic
term can never dominate the dark radiation if
$\tilde{\mu} > 3/(el^2)$.
Conversely, it may dominate for a finite time
if $\tilde{\mu} < 3/(el^2)$ in the range $0 < b < 1$. When the correction
term does dominate, introducing a new variable
$y \equiv (- \ln b)^{1/2}$, implies that the Friedmann
equation can be written as
\begin{equation}
\dot{y} =\frac{\sqrt{3}}{2l} \exp \left[ \frac{3y^2}{2} \right]
\end{equation}
and can therefore be
integrated in terms of the Error function:
\begin{equation}
t = \frac{\sqrt{2\pi}l}{3} {\rm erf} \left[
\left( -\frac{3}{2} \ln b \right)^{1/2} \right] .
\end{equation}

It is of interest to determine whether the logarithmic correction
can result in an epoch of inflationary, accelerated expansion.
A necessary and sufficient condition for inflation is that
$\dot{H} +H^2 >0$. When the logarithmic term
dominates,
it follows from the Friedmann equation (\ref{friedmannf})
that
\begin{equation}
\label{noacc}
\dot{H}+H^2 =-\frac{3}{2l^2b^3} (1-\ln b)
\end{equation}
and since the logarithmic term can only dominate for
$b<1$ (without causing recollapse), Eq. (\ref{noacc})
implies that the expansion rate is always decelerating
and, consequently, the brane does not exhibit inflationary expansion.

Finally, as remarked above,
since $f<0$ for all $b>1$, it follows that the logarithmic
term inevitably comes to dominate the Friedmann equation as the
brane expands, thereby causing the brane to recollapse.
This is radically different to the behaviour of the standard,
spatially flat universe.

In the case of negative spatial curvature, $k=-2$, the logarithmic
term may dominate in the region $b<1$, depending on the choice of
parameter values. The interesting question in this
case, however, is whether the effects of the curvature are
sufficient to prevent a recollapse of the brane.
A necessary and sufficient condition
from Eq. (\ref{friedmannf}) for the brane to expand
to infinity is that
\begin{equation}
\tilde{\mu} + |\tilde{k}| b^2 >  \frac{3}{l^2} b \ln b
\end{equation}
for all $b>1$. Since $\tilde{\mu}>0$, it is
sufficient to show that the stronger constraint
\begin{equation}
\label{sufficient}
b > m \ln  b , \qquad m \equiv 3 \left( \frac{2}{W_3}
\right)^{2/3} \left( \frac{l_P}{l} \right)^{4/3}
\end{equation}
is satisfied, where $l_P^2 =G_4$ is the four--dimensional
Planck length and we have employed Eq. (\ref{tilde}).
If $m<1$,
the gradient of the function $h=b$ always
exceeds the gradient of the function $h=m\ln b$ for $b\ge 1$.
Consequently, since the condition (\ref{sufficient})
is trivially satisfied at $b=1$, it implies that it is always satisfied
for $b>1$. This implies that the curvature term always
dominates the logarithmic term when $b>1$.
Hence, the Hubble parameter never passes through zero, and
recollapse does not take place. Instead,
the brane asymptotes to the Milne universe,
where $a \propto t$ at late times.
It is interesting that this conclusion is independent of
the relative magnitude of the spatial curvature term in the FRW equation
(\ref{friedmannlog}) at any given epoch.
The brane may move an infinite distance from
the bulk black hole, and thereby effectively escape from its
gravitational influence, even if the contribution of
the curvature term to the right--hand--side of
Eq. (\ref{friedmannlog}) is arbitrarily small at very early times $(b \ll
1)$.

The above analysis indicates that the
numerical value of the parameter, $m$, determines
which of the different types of behaviour
for the hyperbolic brane
are physically realistic. A turning point in the
expansion (or contraction) is possible only if $m>1$.
This parameter is specified by the ratio of the
AdS length parameter, $l$, and
the four--dimensional Planck length, $l_P$,
and these parameters can in turn be determined from
the five--dimensional Newton constant, $G_5$, and the bulk
cosmological constant, $\Lambda$. In principle, $m>1$
is therefore possible since it is defined in terms of
the ratio of two independent parameters. However, from the perspective of
the
AdS/CFT correspondence in the large $N$ limit,
it is expected that  $l/l_P\sim N$ and, consequently,
small values for $m$
are favoured from a physical point of view.

\section{Discussion}

In summary, we have
considered the FRW dynamics of a brane propagating
in an AdS bulk space containing a black hole. Taking into
account thermal fluctuations of the black hole entropy and
employing the AdS/CFT
correspondence then leads to the appearance of logarithmic corrections
to the Cardy--Verlinde formula and the associated brane FRW equations.
A qualitative analysis of the role of such logarithmic terms in
brane cosmology was performed and
the regions of parameter space
where the early-- or late--time cosmology is altered were highlighted.
In particular, an open
universe that initially collapses and undergoes a bounce
is allowed, in contrast to the standard FRW cosmology.

We now conclude with a
discussion how these corrections may influence
the stability of AdS black holes.
It is well known that black holes in AdS space undergo a
`Hawking--Page' phase transition when the temperature
of the black hole reaches a critical value \cite{hp}.
The transition between the AdS black hole and pure AdS space occurs because
the former is unstable at low temperatures and the energetically
preferred state corresponds to pure AdS space. At high temperatures,
on the other hand, the black hole is stable and consequently does not decay
into AdS space. This implies that the stability of a specific
AdS space is determined in terms of the corresponding black hole
dynamics.

Such a phase transition has been interpreted by Witten
within the context of the AdS/CFT correspondence in
terms of the confinement--deconfinement transition in the large
$N$--limit of the boundary
super Yang-Mills theory \cite{witten}. Specifically,
deconfinement occurs
when the expectation value of the temporal Wilson loop operator for
the super Yang--Mills theory is non--zero. (This
corresponds to the phase where the AdS black hole is stable).
By contrast, confinement occurs when
this expectation value vanishes (the global AdS space is stable).
Of course, higher--derivative terms originating from
next--to--leading corrections in the large--$N$ expansion
may influence the structure of such an AdS phase
transition \cite{no}. Moreover,
as we now demonstrate, the logarithmic terms
discussed above that arise in
the FRW brane cosmology may also alter
the stability of the AdS black hole and
consequently may introduce new features into
the nature of the Hawking--Page phase transition.

Indeed, let us consider an AdS black hole with a temperature and energy
given by Eq. (\ref{ab4}) and a corrected entropy given by
Eq. (\ref{total}), where ${\cal{S}}_0$ corresponds to the entropy
in Eq. (\ref{ab3}), i.e., ${\cal{S}}_0=W_3a^3_H/(4G_5)$.
The free energy of the black hole with the logarithmic correction
included then follows immediately from the definition, $F\equiv E-T_H
{\cal{S}}$:
\be
\label{correctfree}
F= -{W_3 \over 16\pi G_5} a_{H}^2
\left( {a_{H}^2\over l^2} - 1  \right)
+ {1 \over 2}\left({1 \over 2\pi a_{H}} +{a_{H} \over \pi l^2}
\right)\ln{W_{3} a_H^3 \over 4 G_5}\ ,
\ee
where we have assumed implicitly that $k=2$.
In the absence of any correction term, the free energy
is negative (positive) if $a_{H}^2> l^2$ ($a_{H}^2< l^2$).
Consequently, a large black hole (large $a_H$) is stable
whereas small ones are unstable.
In the case of a large black hole, the
correction term to the free energy in Eq. (\ref{correctfree})
can be neglected. On the other hand,
this term becomes dominant for small $a_H$ and this
implies that for
sufficiently small black holes,
the free energy becomes negative. As a result,
(very) small black holes become stable when the correction term is included.
This conclusion may also hold for small black holes even in a flat
background, where the AdS length parameter,
$l$, tends to infinity. In this case,
small primordial black holes may be stabilized by the
logarithmic correction and this
indicates that the lifetime of
primordial black holes that may have formed in the
early universe would be enhanced relative to that inferred
from a standard analysis.

In view of this, it is important to derive a quantitative
estimate of the role of these logarithmic terms in the
Hawking--Page--Witten
phase transition \cite{witten,mc}.
When $k=2$, regularization of the action when the
black hole solution is substituted follows by subtracting the action of
the AdS vacuum. In the case where $k=0$, however,
it has been argued that it is more straightforward
to subtract the action of the AdS soliton \cite{HM}
instead of the AdS vacuum \cite{SSW}. When $k=0$, the AdS black hole
metric is given by
\begin{eqnarray}
ds_{\rm BH}^2 = - \e^{2\rho_{\rm BH}(r)} dt_{\rm BH}^2
+ \e^{-2\rho_{\rm BH}(r)}dr^2
 + r^2\left(d\phi_{\rm BH}^2 + \sum_{i=1,2}
\left(dx^i\right)^2\right) \nonumber \\
\e^{2\rho_{\rm BH}(r)}={1 \over r^2}\left\{-\mu_{\rm BH}
+ {r^4 \over l^2}\right\}  .
\end{eqnarray}
We assume for simplicity that the three--dimensional Einstein manifold
spanned by the coordinates $\{ \phi_{\rm BH} , x^1, x^2 \}$
is a torus, where
$\phi_{\rm BH}$ has a period of $\eta_{\rm BH}$, i.e.,
we identify $\phi_{\rm BH}\sim \phi_{\rm BH} + \eta_{\rm BH}$.
The AdS soliton solution can then be obtained by exchanging the
signature of $t_{\rm BH}$ and $\phi_{\rm BH}$ such that
$t_{\rm BH}\rightarrow i\phi_s$ and
$\phi_{\rm BH}\rightarrow it_s$.
One then obtains the free energy, $F$, and the entropy, ${\cal S}$,
in the following forms:
\be
\label{FS}
F=F_0\equiv -{\eta_{\rm BH} W_2 l^6 \over \kappa^2}
\left\{\left(\pi T_{\rm BH}\right)^4
 - \left({\pi \over l\eta_{\rm BH}}\right)^4 \right\} \ ,\quad
{\cal S}={\cal S}_0={4\eta_{\rm BH} W_2 l^6 \pi^4 \over \kappa^2}
T_{\rm BH}^3 \ ,
\ee
where $T_{\rm BH}$ is the Hawking temperature of the black hole and
$W_2$ is the volume of the two--torus spanned by $\{ x^i \}$.
Eq. (\ref{FS}) implies that there is a phase transition when
$T_{\rm BH}=( l\eta_{\rm BH})^{-1}$. It then follows that
the black hole is stable
when $T_{\rm BH}>(l\eta_{\rm BH})^{-1}$,  but
unstable when $T_{\rm BH}<(l \eta_{\rm BH})^{-1}$. In
this latter case, the AdS soliton is
the preferred state.

When the logarithmic correction is included and
$F\to F_0 + {1 \over 2}T_{\rm BH}\ln {\cal S}_0$, we find that
\be
\label{FS2}
F= -{\eta_{\rm BH} W_2 l^6 \over \kappa^2}
\left\{\left(\pi T_{\rm BH}\right)^4
 - \left({\pi \over l\eta_{\rm BH}}\right)^4 \right\}
+ {1 \over 2}\ln \left({4\eta_{\rm BH} W_2 l^6 \over \kappa^2}
T_{\rm BH}^3\right) \ .
\ee
Consequently, when $T_{\rm BH}$ is large, the correction does not alter the
behaviour of the free energy, $F$, but when $T_{\rm BH}$ is small,
it becomes dominant and makes the free energy
negative, thereby indicating that the black hole becomes stable at
low temperatures. It would be very interesting to investigate
in more detail the role
of these logarithmic terms in the dual CFT confinement--deconfinement phase
transitions.

\section*{Acknowledgements}

J.E.L. is supported by the Royal Society.
The research by S.N. is supported in part by the Ministry of
Education, Science, Sports and Culture of Japan under
the grant number 13135208.
The  research by S.O. is supported in part by the Japanese Society
for the Promotion of Science under the Postdoctoral Research Programme.

\end{document}